\documentclass[aps,prl,twocolumn,showpacs,groupedaddress]{revtex4-1}
\usepackage{graphicx}
\usepackage{color}
\usepackage{latexsym}
\begin{document}

\preprint{}
\title{Probing locally the onset of slippage at a model multi-contact interface}
\author{V. Romero}
\author{E. Wandersman}
\author{G. Debr\'egeas}
\author{A. Prevost}
\email[]{alexis.prevost@upmc.fr}
\affiliation{CNRS / UPMC Univ Paris 06, FRE 3231, Laboratoire Jean Perrin LJP, F-75005, Paris, France}

\date{\today}

\begin{abstract}
We report on the multi-contact frictional dynamics of model elastomer surfaces rubbed against bare glass slides. The surfaces consist of layers patterned with thousands spherical caps (radius of curvature 100 $\mu$m) distributed both spatially and in height, regularly or randomly. Use of spherical asperities yields circular micro-contacts whose radius is a direct measure of the contact pressure distribution. In addition, optical tracking of individual contacts provides the in-plane deformations of the tangentially loaded interface, yielding the shear force distribution. We then investigate the stick-slip frictional dynamics of a regular hexagonal array. For all stick phases including the initial one, slip precursors are evidenced. They are found to propagate quasi-statically, normally to the iso-pressure contours. A simple quasi-static model relying on the existence of interfacial stress gradients is derived and predicts qualitatively the position of slip precursors.    
\end{abstract}


\maketitle

In recent years, our understanding of the transition from static to dynamic friction has been markedly changed with the development of new imaging techniques to probe spatially the interfacial dynamics at the onset of sliding \cite{RubinsteinNature2004,RubinsteinPRL2007,MaegawaTribLett2010,BenDavidNature2010}. Slip phases were found to involve the propagation of a series of dynamical rupture fronts, far from the classical Amontons-Coulomb's picture. Using true contact area imaging with evanescent illumination of a 1D Plexiglas-Plexiglas plane contact, Fineberg and coauthors \cite{RubinsteinNature2004} measured in particular slow fronts with velocities orders of magnitude lower than the Rayleigh wave velocity, along with sub-Rayleigh and fast intersonic fronts. Slow fronts were also reported to propagate at soft elastomer-roughened glass spherical 2D contacts \cite{AudryEPJE2011} by tracking optically markers positioned on the surface of the elastomer. Br\"orman {\it et al.} \cite{BrormannTribLett2012} extended such studies to micro-structured elastomer surfaces in the form of hexagonal arrays of cylindrical micro-pillars in contact with glass slides, and found again a similar phenomenology. During stick phases, slow slip precursors were also observed well before macroscopic slippage occurs \cite{RubinsteinPRL2007}. In all these experiments, a single physical quantity is measured, either the real area of contact directly related to the local normal stress, or the local interfacial stress using displacement measurements. In a recent work \cite{BenDavidPRL2011}, Ben-David and Fineberg provided both types of measurement in a system treated as a 1D interface. Using an array of strain gauges sensors distributed directly above the interfacial plane, these authors reported strong correlations between the characteristics of the fronts and the ratio of the measured tangential to normal local stresses. For an extended 2D contact, simultaneous measurements of both pressure and tangential interfacial fields is still lacking and out of reach using Ben-David and Fineberg's approach. In addition, it remains unclear what physical mechanism underlies the existence of slip precursors in the stick phase and their velocity, despite numerous theoretical as well as numerical works~\cite{Braun2009,Scheibert2010,DiBartolomeo2010,Scheibert2011,Bouchbinder2011,Amundsen2012,Kammer2012}.

In this Letter, we take advantage of recent developments in micro-milling techniques to design model elastomer multi-contact surfaces \cite{GreenwoodWilliamson1966}. These consist of thousands of spherical caps distributed on top of a rectangular block, all made from the same elastomer. We show that spherical caps provide a unique way to measure optically local normal and shear forces once in contact with bare glass slides. We apply this novel technique to analyze the stick-slip frictional dynamics of an hexagonal array of spherical caps of equal height and radius of curvature. Local analysis first reveals that pressure gradients are inherently present for this plane-plane contact, and second that each slip event is mediated by slip precursors. These are found to be quasi-static and to propagate normally to the iso-pressure lines. We compare our findings with a simplified pressure gradient based model where individual asperities are taken as elastically independent.

\begin{figure}[h]
\center
\includegraphics[width=8cm]{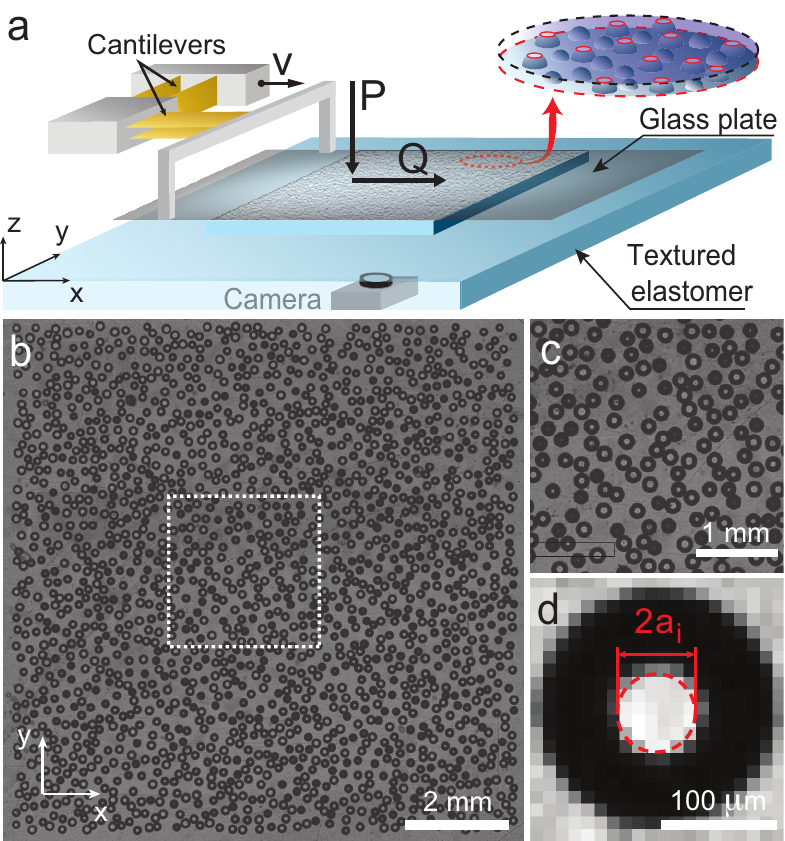}
\caption{(Color online) (a) Sketch of the setup. $P$ and $Q$ are monitored by measuring the deflection of two cantilevers (normal and tangential stiffness \textit{resp.} $810\pm8~\mbox{Nm}^{-1}$ and $10673\pm285~\mbox{Nm}^{-1}$) with capacitive position sensors (MCC-20 and MCC-10, Fogale Nanotech). (b) Image of a RR sample ($\Phi=40\%$) in contact with a glass slide. (c) Close-up on the dashed line delimited zone in (b). (c) Single asperity in contact (contact diameter $2 a_i$).}
\label{Fig1}
\end{figure}

Micro-structured surfaces are made of a crosslinked PolyDimethylSiloxane (PDMS Sylgard 184, Dow Corning) cured for 48 hours in an oven at 75$\,^{\circ}\mathrm{C}$. They are obtained by pouring a PDMS melt-crosslinker liquid mixture in a 10:1 mass ratio in a Plexiglas mold fabricated with a desktop CNC Mini-Mill machine (Minitech Machinary Corp., USA). The molds consist of $10 \times 10~$mm$^2$ square cavities, $2.5~$mm deep. Their bottom surface is covered with spherical holes whose constant radius of curvature $R=100~\mu$m is directly set by the ball miller used. Holes are positioned spatially with $1~\mu$m resolution either over a regular lattice or at random and their maximum depths are either equal or taken at random from a uniform distribution in the range 40-60 $\mu$m. Resulting PDMS surfaces are decorated with spherical caps which match the designed pattern. For the present work, different types of patterns were fabricated -- two hexagonal lattices with a base surface coverage $\Phi=0.4$, one with constant height asperities (LC) and one with random height asperities (LR), and two random distributions with random height asperities (RR), with $\Phi=0.2$ and 0.4. Samples are maintained by adhesion against a solid glass plate and put in contact with a clean bare glass slide under constant normal load $P$. The glass slide is mounted on a double cantilever system which allows to measure both $P$ and the applied shear force $Q$ with mN resolution in the range [0--2.5] N (Fig.~\ref{Fig1}a, \cite{PrevostEPJE2013}). The latter is attached to a motorized stage (LTA-HL, Newport) which can be driven at constant velocity $v$ in the range [4--1000] $\mu$m/s. A white light LED array illuminates the interface in transmission through the glass slide. On the opposite side, a CMOS sensor based camera (Photon Focus, $1380\times1024~$ pixels$^2$, 8 bits, 30 Hz at full frame) makes an image of the interface. In addition, a fast camera (Photron Fastcam APX-RS, $1024\times1024$ pixels$^2$, 8 bits) operating at 1000 Hz was used to probe the fast dynamics during stick-slip events. As shown on Figs.~\ref{Fig1}b-c-d, light is transmitted at every single micro-contact and refracted by the spherical caps elsewhere, resulting in a myriad of white circular spots, whose radii $a_i$ can be extracted using standard image analysis techniques (Fig.~\ref{Fig1}d).

\begin{figure}[h]
\includegraphics[width=\columnwidth]{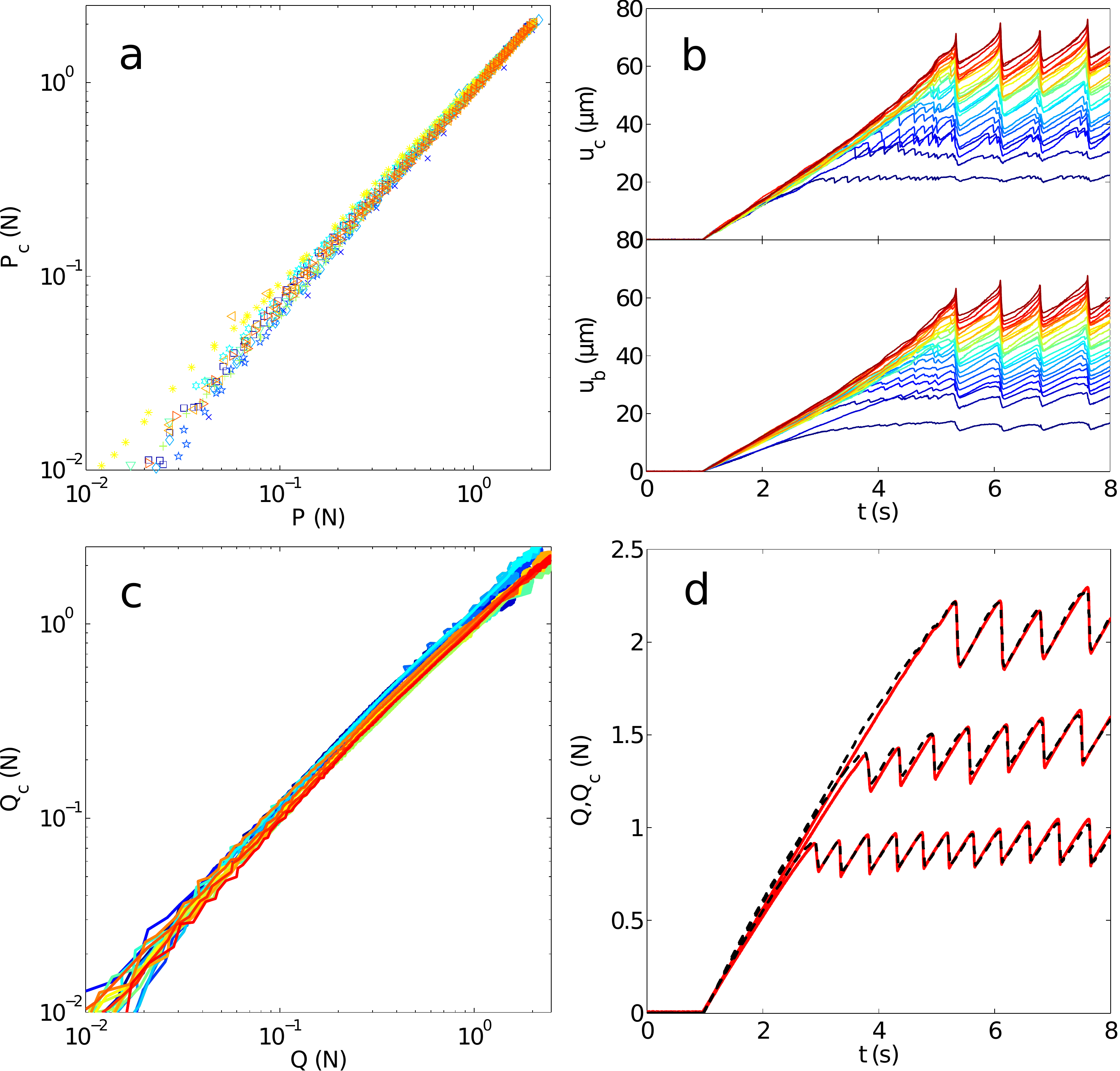}
\caption{(Color online) (a) $P_c $ \textit{vs} $P$ for all patterns (different colored symbols) loaded normally. (b) Micro-contacts (\textit{resp.} back layer) displacements $u_c(t)$ (\textit{resp.} $u_b(t)$) for 23 micro-contacts chosen at random in the LC sample ($v=80~\mu$m/s, $P=2$~N). $p_i$ increases from bottom to top (blue to red). (c) $Q_c$ \textit{vs} $Q$ for all patterns (different colored lines) in shear experiments. (d) $Q(t)$ (solid lines) and $Q_c(t)$ (dashed lines) for the LC pattern with $P = {0.5,1,2}$~N (bottom to top) and $v=80~\mu$m/s.}
\label{Fig2}
\end{figure}

\indent Assuming Hertz's model to describe the contact mechanics between the spherical caps and the glass slide, the local applied load $p_i$ is given by 

\begin{equation}
p_i=\frac{4 E a_i^3}{3 (1-\nu^2) R} 
\label{EqHertz}	
\end{equation}

\noindent where $E$ is the elastomer Young's modulus and $\nu$ its Poisson's ratio taken as 0.5 \cite{PoissonRatio}. This allows computing the total normal load $P_c=\sum_{i} p_i$. For all experiments, a linear relationship is systematically found between $P_c$ and $P$ over two orders of magnitude in $P$ in the range [0--2.5]~N, irrespective of the type of disorder and pressure distributions (Fig.~\ref{Fig2}a). Hertz assumption is thus clearly validated in normal contact conditions. However, the slope of $P_c$ versus $P$ depends slightly on the optical threshold used to detect $a_i$. To recover a unit slope, we thus calibrated the optical threshold with a reference sample whose Young's modulus $E = 4.1 \pm 0.1~$MPa has been measured independently with a JKR test \cite{Wu-BavouzetPRE2010}. We then kept the resulting threshold for other samples and tuned $E$ within experimental errors to recover a unit slope. Upon shearing the interface, obtained by driving the translation stage at constant velocity $v$ in the range [$20~\mu$m/s--$120~\mu$m/s], the micro-contacts size changes marginally from circular to slightly elliptic, still allowing local normal loads to be extracted within Hertz model's assumption.

\indent Contrary to the usual pillar geometry of asperities \cite{Wu-BavouzetPRE2010,MurarashSoft2011,DegrandiFaradDiscuss2012,BrormannTribLett2012}, spherical asperities do not bend nor buckle. It is thus possible to locate unambiguously with sub-pixel accuracy ($1/24~$ pixels, $\sim 400~$nm) positions of the micro-contacts centers and follow, using a custom made algorithm written in Matlab (MathWorks), their displacements with respect to their initial position, $u_c$ (Fig.~\ref{Fig2}b, upper panel). The same methods allow to extract the displacement of the back layer by monitoring positions of the base of spherical asperities, $u_b$ (Fig.~\ref{Fig2}b, lower panel).  Defining $\delta=u_c-u_b$ as the displacement of the cap top with respect to the back layer, we measured $\delta \approx \alpha v t$ with $\alpha \approx 0.032$. Neglecting any micro-slip at the edges of the micro-contacts \cite{Cattaneo-RANL-1938,Mindlin-JAppMech-1949,ChateauminoisPRE2010,PrevostEPJE2013}, one can show that the shear force $q_i$ acting on an asperity is proportional to $a_i$ \cite{JohnsonBook2003}, as

\begin{equation}
q_i =  \frac{8 E a_i}{3(2-\nu)}~\delta
\label{EqQ}
\end{equation}

The total shear force $Q_c$ can then be computed writing that $Q_c = \sum_{i} q_i$. For all patterns and experiments, Eq.~\ref{EqQ} provides a good approximation for the local shear force as shown on Figs.~\ref{Fig2}c and ~\ref{Fig2}d. A one-to-one linear relationship between $Q_c$ and $Q$ over two orders of magnitude is found (Fig.~\ref{Fig2}c). Fig.~\ref{Fig2}d illustrates this agreement on the example of the LC pattern with $Q(t)$ and $Q_c(t)$ at different $P$.

\begin{figure}[h]
\includegraphics[width=8cm]{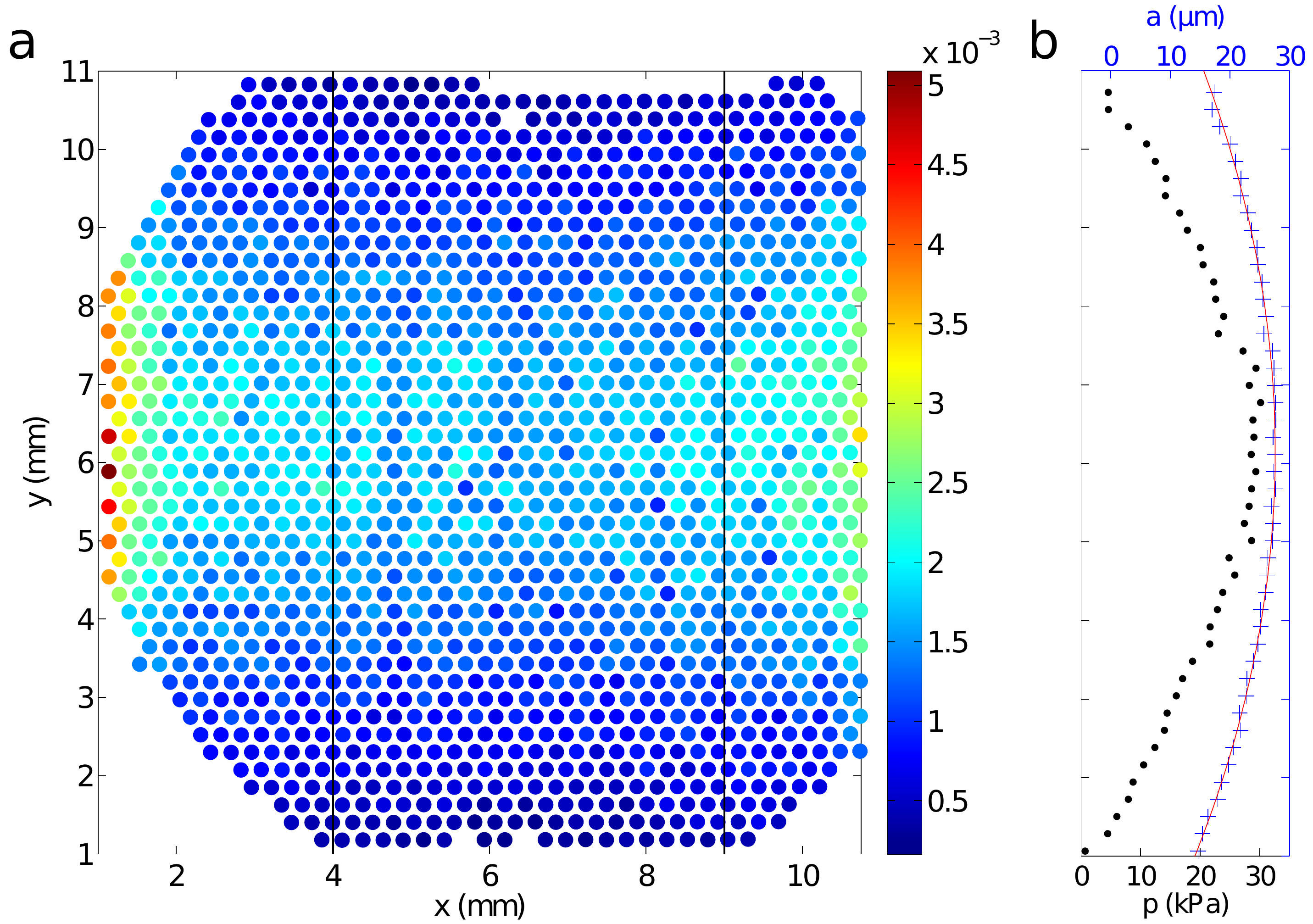}
\caption{(Color online) (a) Spatial distribution of normal local forces (in N) for the LC pattern in the stick-slip regime ($P=2.36~$N). (b) Pressure ($\bullet$) and radii (+) distributions averaged along $x$ in the region bounded by the two vertical lines in (a). The line is a fit $a(y)=a_0+ a_1~y + a_2~y^2$ with $\{a_0,a_1,a_2\}=\{8.37~\mu$m, 6.27~10$^{-3}$, -0.51~m$^{-1}\}$.}
\label{Fig3}
\end{figure}

\begin{figure}[h]
\includegraphics[width=\columnwidth]{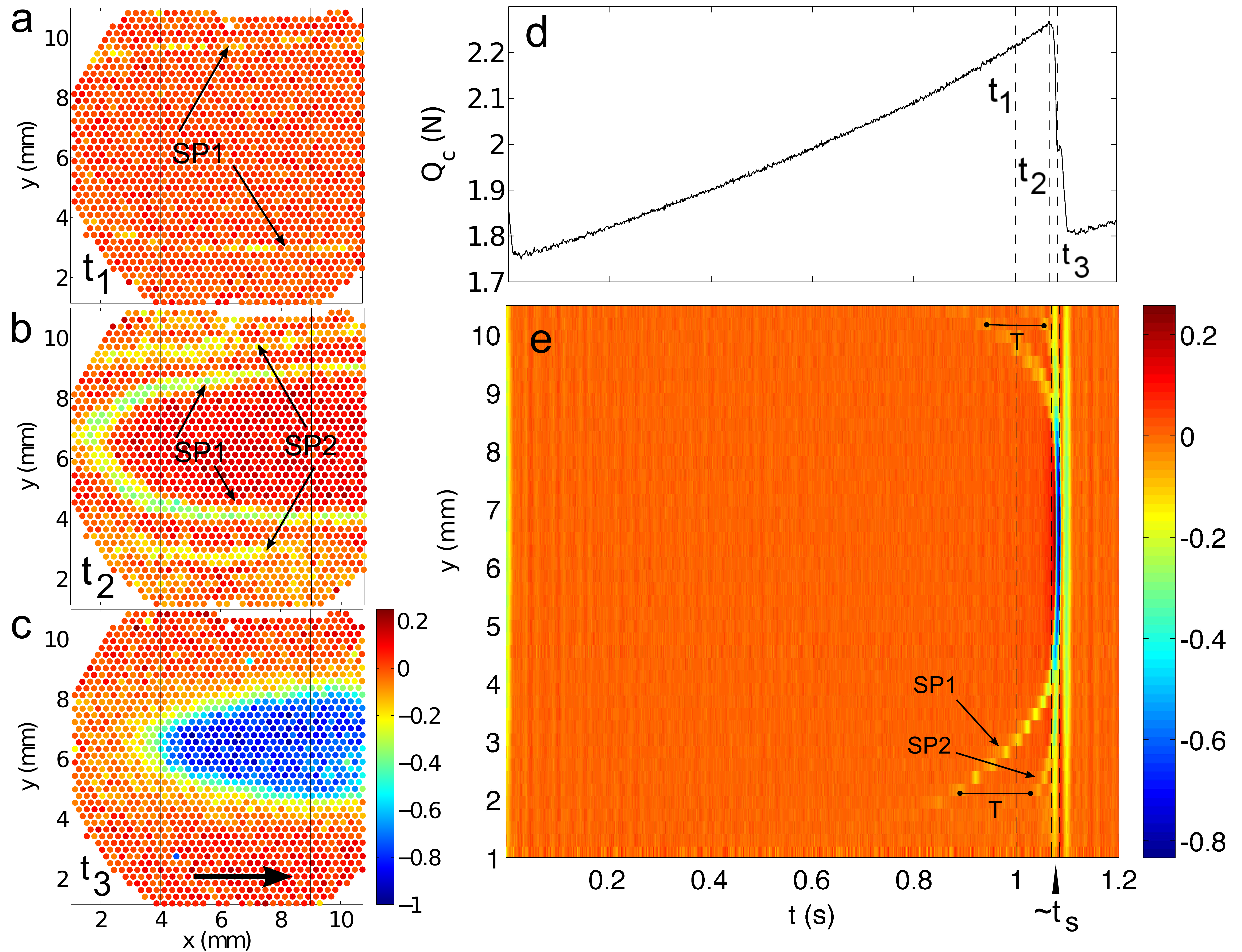}
\caption{(Color online) (a) Velocity field snapshot at time $t_1$. SP1 stands for 1$^{st}$ slip pulse. (b) Same at time $t_2$. SP2 stands for 2$^{nd}$ slip pulse. (c) Same at time $t_3$. The black arrow shows the direction of sliding. On (a)-(b)-(c), the vertical lines delimit the central region defined in Fig.~\ref{Fig3}a. (d) $Q_c$ curve \textit{vs} time for the stick-slip event of (a), (b) and (c). Dashed lines are drawn at times $t_1, t_2$ and $t_3$. (e) Spatiotemporal plot of the velocity field along $y$ averaged for 4 mm $\leq x \leq$ 9 mm. Velocities are given in mm/s. SP2 starts propagating with a delay $T$ with respect to SP1's initial trigger event.}
\label{Fig4}
\end{figure}

We now turn onto analyzing in details the frictional dynamics of the LC pattern, which is the simplest available texture, when it is sheared against a glass plate along $x$. For that system, $Q$ is found to increase up to a static threshold $Q_s$, beyond which a stick-slip instability always sets in at all normal loads and within the tested driving velocity range of [$20~\mu$m/s--$120~\mu$m/s] (Fig.~\ref{Fig2}d). The spatial distribution of local normal forces is found to be non-uniform with a characteristic saddle-like shape (Fig. \ref{Fig3}a). As $Q$ increases, the pressure distribution evolves continuously from an initial state at $Q = 0$ (not shown) to a slightly different one at $Q = Q_s$, essentially characterized by lower normal forces at small $x$. Beyond $Q_s$, the distribution (Fig.~\ref{Fig3}a) is time invariant. Such non-uniformity presumably results from combined effects of the existence of a curvature of the sample at long wave lengths, contact loading history and Poisson expansion \cite{BrormannTribLett2012}. Analysis of the displacement curves $u_c(t)$ reveals that during initial and subsequent stick phases, slip precursors nucleate and eventually invade the whole contact. In the stick-slip regime, they can be best evidenced when looking at velocity field snapshots, obtained by deriving the 2D displacement fields $u_c$ with respect to time (Figs.~\ref{Fig4}a-b-c at three different instants during the stick-slip event of Fig. \ref{Fig4}d). In the stick phase ($t \leq t_s$, where $t_s$ is the time of slip, different for each event), they appear as spatially localized structures with large negative velocities, indicative of a collective back-snapping of the micro-contacts (Figs.~\ref{Fig4}a and ~\ref{Fig4}b). A secondary slip pulse also forms several asperities behind the first one (Fig.~\ref{Fig4}b). In the slip phase ($t > t_s$) however, all remaining micro-contacts back-snap coherently. These two consecutive slip pulses are systematically observed for all stick-slip events, and are always found to nucleate on the edges of the contact. When focusing on the central band $4 \leq x \leq 9~$mm, front lines are essentially oriented along $x$ normally to the iso-pressure lines (\textit{see} Fig.~\ref{Fig4}a) \cite{footnote}. Within this band, the velocity field along the $y$ direction is averaged over $x$ to help visualizing how the front propagates spatially over time.

On the resulting spatiotemporal plot (Fig.~\ref{Fig4}e), both first and second slip pulses are visible, each of them consisting of two branches, almost symmetric with respect to the $y \approx 6~$mm axis. The first slip pulse appears to propagate initially with a constant velocity before continuously accelerating as $t$ approaches $t_s$, reaching a maximum velocity of about 10 mm/s, three orders of magnitude lower than the Rayleigh wave velocity ($\approx$ 10 m/s for this elastomer). The observed scenario remains qualitatively similar for the first loading stick phase, but slip precursors are more heterogeneously distributed, preventing a direct quantitative analysis. For the present work, we have thus chosen to focus on the stick-slip regime only.

\begin{figure}[h]
\includegraphics[width=\columnwidth]{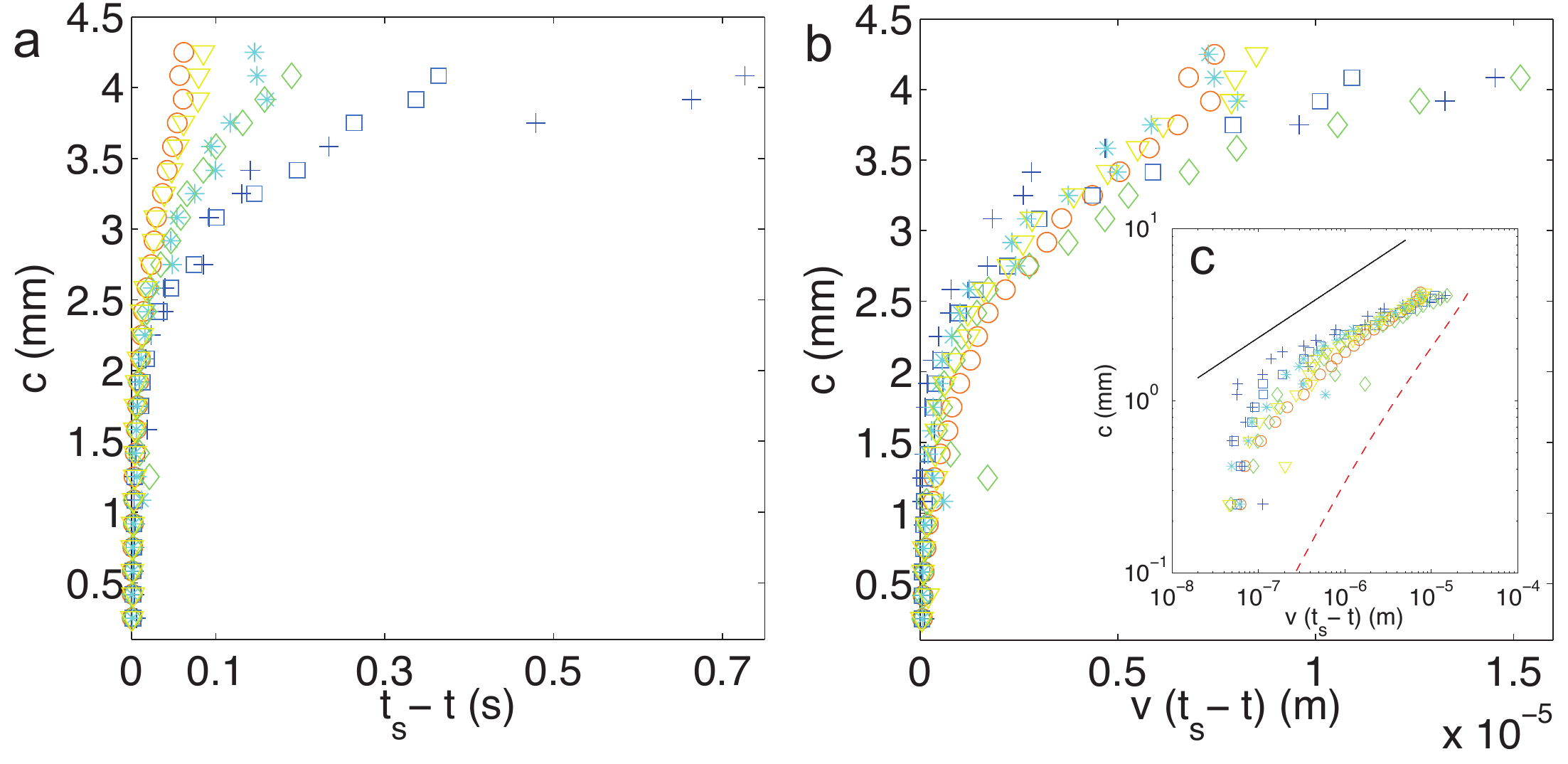}
\caption{(Color online) (a) $c$ \textit{vs.} $(t_s-t)$ for the first slip pulse and $v = 20 (+), 30 (\Box),  50 (\diamond),  80 (\ast), 100 (\bigtriangledown), 120 (\circ)~\mu$m/s. (b) $c$ \textit{vs.} $v (t_s-t)$. (c) Log-log plot of (b). The solid line is a power law of exponent $1/3$. The dashed line is the model prediction.}
\label{Fig5}
\end{figure}

For each stick-slip event, front positions were obtained by detecting individual times of slip for each asperity in contact, using their displacement $u_c(t)$, allowing to obtain them with a better accuracy. Mean front positions versus mean times of slip were deduced by averaging both individual slip times of all asperities at the same $y$-position (within the central $x$-band) and mean front positions on all stick-slip events. Similarly to the velocity spatiotemporal representation, such curves are almost axisymmetric around $y \approx 6$~mm, allowing to extract the distance $c$ to this axis of symmetry, which is a direct measure of the remaining stick zone extension. This procedure was applied for 6 experiments at $P=2.36~$N with increasing driving velocities $v$. Figure \ref{Fig5} shows the resulting $c$ \textit{vs} $(t_s-t)$ for the first slip pulse (Fig.~\ref{Fig5}a) and the same data with the time axis multiplied by $v$ (Fig.~\ref{Fig5}b). All curves at different $v$ are found to overlap on the same master curve, suggesting that propagation of slip precursors results from a quasi-static mechanism. 

Quasi-static slip precursors have already been reported numerically \cite{Scheibert2010,Otsuki2013,TaloniArXiv2013} but not experimentally, and its underlying physics remains elusive. In an attempt to provide an answer within the framework of our measurements, let us model our system with asperities distributed along $y$ (the direction normal to the iso-pressure lines) on a unidimensional regular lattice of lattice constant $b$, and let us neglect the elastic interaction between them (\textit{i.e.} absence of any back layer). In the classic Amontons-Coulomb's friction description, slip of an asperity $i$ occurs once the local shear load $q_i$ satisfies $q_i = \mu_s p_i$, where $\mu_s$ is a static friction coefficient. Combining Eqs.~\ref{EqHertz} and \ref{EqQ} yields the maximum displacement $\delta_s^i$ beyond which slip occurs and the asperity snaps back, as

\begin{equation}
\delta_s^i=\frac{\mu_s a_i^2}{R}
\label{thsld}
\end{equation}

An asperity $i$ initially at position $y_0^i=b\times i$ will slip when its position reaches $y_s^i=y_0^i+\delta_s^i\approx y_0^i$, since  $\delta_s^i \ll b$. Combined with Eq.~\ref{thsld}, this expression allows to predict for a given pressure profile, the front position and time of slip, respectively $y_w=y_s^i$ and $t_s^i=\delta_s^i/(\alpha v)$. In an ideal plane-plane contact where pressure is uniformly distributed over the contact (to the exception of the edges of the contact), all asperities should slip simultaneously and no slip pulse should be observed. In our experiments however, pressure gradients are clearly present along $y$ as evidenced on the example of Fig.~\ref{Fig3}a. Taking a continuous limit, the contact radius $a_i$ \textit{vs} position in the sample can be reasonably well fitted by a parabola $a(y) = a_0+ a_1~y + a_2~y^2$ (\textit{see} Fig.~\ref{Fig3}b). Using this expression with Eq.~\ref{thsld} provides directly the position of the front with respect to its position at threshold, $c(\delta)=y_w(\delta_s)-y_w(\delta)$, where $\delta_s=\frac{\mu_s}{R}\left(a_0 - \frac{a_1^2}{4 a_2}\right)^2$ is the threshold displacement at $t=t_s$. It reads

\begin{equation}
c(\delta)=-\frac{1}{2 a_2}\sqrt{a_1^2-4a_2\left(a_0-\sqrt{\frac{R}{\mu_s}\delta}\right)}
\label{xwdelta}
\end{equation}

This quasi-static model can be extended to any pressure distribution if needed, and provides a description of the first loading phase, where all micro-spheres start from their initial unloaded position. Once a sphere slips, it relaxes back from its maximum displacement $\delta_s^i$ by $\delta_r^i = \frac{\Delta\mu}{\mu_s} \delta_s^i$ before the beginning of a next loading phase, where $\Delta\mu = \mu_s-\mu_d$ with $\mu_d$ a dynamical friction coefficient. The model can be extended to the stick-slip events by replacing $\mu_s$ by $\Delta \mu$ in Eq.~\ref{xwdelta}. Note that close to the threshold ($\delta-\delta_s \ll\delta_s$), $c(\delta)$ behaves asymptotically as

\begin{equation}
c(\delta) \sim \sqrt{\frac{2R(\delta_s-\delta)}{\Delta \mu (a_1^2 - 4 a_0 a_2)}}
\label{Asy}
\end{equation}  

\noindent and one thus expects $c(\delta)$ to follow a power law of exponent 1/2. Predictions of Eq.~\ref{xwdelta} are plotted on Fig.~\ref{Fig5}c, with $\{a_0, a_1,a_2\}$ given by the parabolic fit (\textit{see} caption of Fig.~\ref{Fig3}b) and $\Delta \mu = 0.157$, obtained by averaging values of $\Delta \mu$ for all experiments. The predicted curve qualitatively succeeds in reproducing the measured trend and right order of magnitude of $c(\delta)$, but fails quantitatively, as measured $c$ values are systematically above it. In addition, careful examination of the data tend to suggest that $c$ follows indeed a power law, but with a characteristic exponent closer to 1/3 than 1/2 (Fig.~\ref{Fig5}c). The present toy model lacks several ingredients which could explain the observed discrepancies. First, it is limited to a 1D description whereas the slip propagation is clearly 2D. Second, it does not take into account the elastic coupling between neighboring asperities connected to the elastic back layer. Including both effects within a full contact mechanics approach is expected to yield an improved quantitative comparison, which is beyond the scope of the present Letter.\\
\indent Beyond its obvious limitations, the present toy model provides however a simple zeroth order mechanism to generate slip pulses, based on the existence of interfacial stress gradients. Interestingly, it also allows to give an explanation for the existence of second slip pulses whose propagation is delayed by $T$ with respect to the first slip pulse, as evidenced on Fig.~\ref{Fig4}e. This delay results from the addition of (i) the individual relaxation time $\tau$ of a given sphere sliding back from its maximum position $\delta_s^i$ of the distance $\delta_r^i$, plus (ii) the time to reach $\delta_s^i$ again. As a result, one expects $T=\tau+\delta_r^i/(\alpha v)$. Such velocity dependence of $T$ is actually verified experimentally (not shown), asserting furthermore the quasi-static character of the measured slip pulses. Taking $\tau=7.6~\pm~0.5$ ms, obtained by averaging times of relaxation for all individual trajectories, one gets $\delta_r^i \approx 0.35~\mu$m, to be compared to the measured averaged value of $1~\mu$m. In addition, the second slip pulse can only be identified if $T(i=0)< t_s$, duration of the slip event for which the global collapse of the interface happens. This criterion thus gives a limiting driving velocity $v_l$ above which no second slip pulse can be observed, 
$v_l=\frac{1}{\tau}(\delta_r-\delta_r^{i=0})=\frac{\Delta\mu}{\alpha R \tau}((\frac{4 a_2 a_0 - a_1^2}{4a_2})^2-a_0^2)$.
Using the experimental values and $\tau=7.6$ ms, one gets $v_l \approx 4.4$~mm/s. For $v > v_l$ the relaxation time of the spheres are long in comparison to the wave propagation time, and the propagation of the $n+1^{th}$ wave will start whereas the relaxation of the spheres is not finished, leading to a blur of the force signal. This is not observed in our experiments since the maximum tested driving velocity ($v=0.12$~mm/s) remains small compared to $v_l$.\\
\indent The present results have been purposely limited to the stick-slip regime where slip precursors have been clearly identified and could be characterized and compared to a simple non-interacting model. As mentioned earlier, a similar phenomenology is observed for the first stick event, and will be explored in more details in a future work. Our results demonstrate how the combination of surface micro-patterning and interface imaging allows one to access the micro-mechanics at the level of single asperities. This has been applied to the case of a hexagonal array of equal height micro-asperities, revealing that slip precursors propagate quasi-statically orthogonally to the iso-pressure lines. It will be extended to more elaborate patterns in a future work.

\begin{acknowledgments}
We acknowledge funding from ANR (DYNALO NT09-499845). We also thank A. Chateauminois and C. Fr\'etigny (PPMD, ESPCI, France) for fruitful discussions, and are indebted to R.~Candelier for his help in designing at an early stage the molds used in these experiments. V.~R. is also grateful for CONICYT financial support.
\end{acknowledgments}


\begin{thebibliography}{27}
\bibitem{footnote} The latter observation was systematically reproduced with samples prepared with the same hexagonal pattern but which produced a different pressure spatial distribution once in contact with the glass slide.
\end{thebibliography}
\end{document}